\documentclass[runningheads]{llncs}

\usepackage[american]{babel}
\usepackage[english=american]{csquotes}

\usepackage[shortcuts]{glossaries}
\glsdisablehyper
\PassOptionsToPackage{hyphens}{url}
\usepackage{hyperref}
\usepackage{graphicx}
\usepackage{subcaption}
\usepackage{xcolor}
\usepackage{caption}
\usepackage{subcaption}
\usepackage{listings}
\usepackage{paralist}
\usepackage{enumitem}
\usepackage{amssymb}
\usepackage{pifont}
\usepackage{xspace}
\usepackage{comment}
\usepackage{fancyvrb}
\usepackage{graphicx,txfonts}
\usepackage{multirow}
\usepackage{float}
\usepackage{multicol}
\usepackage{verbatim}
\usepackage[utf8]{inputenc}
\usepackage[T1]{fontenc}
\usepackage{framed}
\usepackage{setspace}
\usepackage{cleveref}
\usepackage[htt]{hyphenat}
\usepackage[textsize=footnotesize, textwidth=40mm]{todonotes}
\usepackage{booktabs}
\usepackage{tcolorbox}
\usepackage{color, colortbl}
\usepackage{threeparttable}
\usepackage{makecell}
\usepackage{rotating}
\presetkeys{todonotes}{fancyline}{}
\renewcommand{\texttt}[1]{%
  \begingroup
  \ttfamily
  \begingroup\lccode`~=`/\lowercase{\endgroup\def~}{/\discretionary{}{}{}}%
  \begingroup\lccode`~=`[\lowercase{\endgroup\def~}{[\discretionary{}{}{}}%
  \begingroup\lccode`~=`.\lowercase{\endgroup\def~}{.\discretionary{}{}{}}%
  \catcode`/=\active\catcode`[=\active\catcode`.=\active
  \scantokens{#1\noexpand}%
  \endgroup
}

\newif\ifanonymous
\newif\ifnotanonymous

\ifdefined\isanonymous
\anonymoustrue
\fi

\ifanonymous\notanonymousfalse
\else\notanonymoustrue
\fi

\newcommand{\splitcell}[1]{
  \begingroup
  \renewcommand{\arraystretch}{1}%
  \begin{tabular}{@{}c@{}}#1\end{tabular}%
  \endgroup
}

\newcommand{\dOne}{\ding{182}\xspace}
\newcommand{\dTwo}{\ding{183}\xspace}
\newcommand{\dThree}{\ding{184}\xspace}

\newcommand{\dCOne}{\ding{192}\xspace}
\newcommand{\dCTwo}{\ding{193}\xspace}
\newcommand{\dCThree}{\ding{194}\xspace}
\newcommand{\dCFour}{\ding{195}\xspace}

\newcommand{\cmark}{\ding{51}}%
\newcommand{\xmark}{\ding{55}}%

\newacronym{abi}{ABI}{application binary interface}
\newacronym{api}{ABI}{application programming interface}
\newacronym{aslr}{ASLR}{address space layout randomization}
\newacronym{bss}{BSS}{block starting symbol}
\newacronym{cfi}{CFI}{Control-Flow Integrity}
\newacronym{cet}{CET}{Control-flow Enforcement Technology}
\newacronym{cve}{CVE}{Common Vulnerability Enumeration}
\newacronym{cwe}{CWE}{Common Weakness Enumeration}
\newacronym{gcs}{GCS}{Guarded Control Stack}
\newacronym{glibc}{glibc}{GNU C library}
\newacronym{ebp}{\protect\texttt{\%ebp}}{x86 base pointer register}
\newacronym{isa}{ISA}{instruction set architecture}
\newacronym{pa}{PA}{Pointer Authentication}
\newacronym{pac}{PAC}{pointer authentication code}
\newacronym{ram}{RAM}{random access memory}
\newacronym{rbp}{\protect\texttt{\%rbp}}{base pointer register}
\newacronym{rsp}{\protect\texttt{\%rsp}}{stack pointer register}
\newacronym{rop}{ROP}{return-oriented programming}
\newacronym{smt}{SMT}{simultaneous multithreading}
\newacronym{os}{OS}{operating system}
\newacronym{nist}{NIST}{National Institute of Science and Technology}
\newacronym{x86-64}{x86-64}{64-bit x86}

\makeatletter
\def\bstctlcite{\@ifnextchar[{\@bstctlcite}{\@bstctlcite[@auxout]}}
\def\@bstctlcite[#1]#2{\@bsphack
  \@for\@citeb:=#2\do{%
    \edef\@citeb{\expandafter\@firstofone\@citeb}%
    \if@filesw\immediate\write\csname #1\endcsname{\string\citation{\@citeb}}\fi}%
  \@esphack}
\makeatother

\begin{document}

\bstctlcite{lncs:BSTcontrol}

\title{Do we still need canaries in the coal mine? Measuring shadow stack effectiveness in countering stack smashing}
\titlerunning{Measuring shadow stack effectiveness in countering stack smashing}
\ifanonymous
\author{}
\institute{}
\fi
\ifnotanonymous
\author{Hugo Depuydt\inst{1} \and
Merve G\"ulmez\inst{2} \and
Thomas Nyman\inst{3} \and Jan Tobias M\"uhlberg\inst{4}}
\authorrunning{Depuydt et al.}
\institute{ENS Rennes, France \email{hugo.depuydt@ens-rennes.fr} \and
Ericsson Security Research, Sweden \email{merve.gulmez@ericsson.com} \and
Ericsson Product Security, Sweden \email{thomas.nyman@ericsson.com} \and
Universit\'e Libre de Bruxelles, Belgium \email{jan.tobias.muehlberg@ulb.be}}
\fi
\maketitle

\begin{abstract}
Stack canaries and shadow stacks are widely deployed mitigations to
memory-safety vulnerabilities. While stack canaries are introduced by the
compiler and rely on sentry values placed between variables and control
data, shadow stack implementations protect return addresses explicitly and
rely on hardware features available in modern processor designs for
efficiency. In this paper we hypothesize that stack canaries and shadow
stacks provide similar levels of protections against sequential stack-based overflows.
Based on the Juliet test suite, we evaluate whether \gls{x86-64}
systems benefit from enabling stack canaries in addition to the \gls{x86-64} shadow
stack enforcement. We observe divergence in overflow detection rates
between the GCC and Clang compilers and across optimization levels, which we attribute to differences in stack
layouts generated by the compilers. We also find that \gls{x86-64} shadow stack
implementations are more effective and outperform stack canaries when combined with a
stack-protector-like stack layout.
We implement and evaluate an  enhancement to the Clang \gls{x86-64} shadow stack
instrumentation that improves the shadow stack detection accuracy based on
this observation.

\end{abstract}

\section{Introduction}

The urgency of mitigating memory-safety vulnerabilities in software developed with the C and C++ programming languages has grown under increasing regulatory scrutiny~\cite{ONCD24}.
Memory-safety issues are one of the oldest problems in computer security and remain a persistent challenge despite decades of advancements in both offensive and defensive techniques~\cite{Szekeres13}.
Among these, stack canaries~\cite{Cowan98} stand out as one of the earliest systematic mitigations to achieve widespread adoption.
In this paper, we reassess stack canaries in light of modern hardware-assisted mitigations, particularly shadow stacks~\cite{Burow19}, now operational in commodity \gls{x86-64} systems~\cite{Larabel23,Larabel24}.

Stack canaries---a reference to the historic practice of bringing canary birds into coal mines as they would be affected by toxic gases earlier than the miners---are sentinel values placed between local variables and control data on the stack to detect buffer overflows.
Shadow stacks, in contrast, specifically protect function return addresses, preventing exploits such as \gls{rop}~\cite{Shacham07} that hijack a program's control flow.
While shadow stacks target a different threat model, both techniques defend against sequential overflows that corrupt the stack canary or return address.
We hypothesize that, with modern compiler optimizations omitting other control data from stack frames, stack canaries and shadow stacks offer comparable protection against sequential overflows.

\noindent
\textbf{This paper and contributions.}
This paper investigates whether conventional stack canaries still offer security benefits when paired with shadow stack enforcement.
We evaluate the effectiveness of both techniques on modern \gls{x86-64} systems using the \acrshort{nist} Juliet C/C++ Test Suite~\cite{Boland12} which contains a wide range of C/C++ code examples with buffer-overflows among the 118 \gls{cwe} categories the suite covers.
Our key contributions and findings include:

\begin{enumerate}
  \item \textbf{Systematic evaluation between GCC and Clang.} We evaluate the
  effectiveness and performance of stack canaries and \gls{x86-64} shadow stack
  in GCC and Clang and show differences in the detection accuracy between the
  compilers. Overall, Clang demonstrates a better detection rate with stack
  canaries than GCC, while shadow stacks alone detect significantly less buffer
  overflows in the sample set compared to stack canaries. We further investigate
  the reasons for this difference.
  \item \textbf{Impact of compiler stack layouts.} The stack layout generated by the compiler has a significant impact on detection accuracy for both stack canaries and shadow stack. The stack layout varies between the different compilers, the level of program optimizations used, and between different variants of the stack-canary instrumentation, i.e., the different option variants in the \texttt{-fstack-protector} family.
  \item \textbf{Enhancements to Clang's shadow stack support.} To enhance the protection the \gls{x86-64} shadow stack offers against sequential buffer overflows, we propose new Clang compiler options that emulate stack-protector layouts while relying on shadow stack checks. Our evaluation shows these new options improve detection accuracy while allowing stack canary checks to be omitted and incur only a small ($\approx$~0.8\% and $\approx$~0.25\% on rate and speed test suites respectively) performance degradation which is lower than that of the corresponding stack canaries ($\approx$~2.18\% and $\approx$~3.21\%, when applied to all functions) and comparable to that of conventional \gls{x86-64} shadow stacks ($\approx$~0.99\% and $\approx0.40$).
\end{enumerate}

Our observations have already been shared with security researchers in the GCC and Clang communities, with whom we confirmed that our findings can be publicly disclosed.

\section{Background}\label{sec:background}

Over half a century since their discovery~\cite{Anderson72}, memory-safety vulnerabilities have become the most prevalent class of software vulnerability~\cite{ONCD24}.
Major software manufacturers, such as Microsoft and Google~\cite{Google21}, attribute up to 70\% of vulnerabilities discovered in their products to memory-safety issues~\cite{MSRC19,Google21}.
Examples of vulnerabilities, attacks, and outages attributed to memory-safety issues include the Heartbleed bug in OpenSSL~\cite{Synopsys14}, the BLASTPASS exploit chain used to deliver commercial spyware~\cite{CitizenLab23}, and the CrowdStrike outage of 2024~\cite{CrowdStrike24}.
The cost to businesses, and society as a whole, of responding to cyber
emergencies caused by memory-safety bugs have prompted cybersecurity authorities
to urge software manufacturers to adopt memory-safe programming languages,
memory-safe hardware, and develop empirical metrics to measure
``\emph{cybersecurity quality}''~\cite{ONCD24}.
Nevertheless, the massive scale at which C and C++ are deployed across the software industry today means memory-unsafe code will remain for the foreseeable future~\cite{Google24}.

Code written in, C, C++, or any other language that is compiled down to native code can, however, be hardened against memory-safety bugs, such as buffer overflows, through the use of compiler-inserted mitigations~\cite{OpenSSFcontributors24}. These are designed to render residual memory-safety vulnerabilities non-exploitable by forcefully terminating a vulnerable application if a bug is triggered.
However, compiler developers lack good tools to verify whether such security hardening features generate code correctly~\cite{Beyls24}.
This is because conventional software testing practices only verify whether hardened code produces the expected output for a given input; not whether the hardening feature operates correctly across different code patterns when a memory-safety bug is triggered.
To address this gap, we perform an empirical evaluation of two production-grade security hardening features available in the GCC and Clang open-source compilers focusing on their effectiveness in detecting classic \emph{buffer overflows} under a variety of scenarios.

\subsection{Buffer overflows}

\begin{figure*}[t!]
    \centering
    \begin{subfigure}[t]{0.49\textwidth}
        \includegraphics[width=\linewidth]{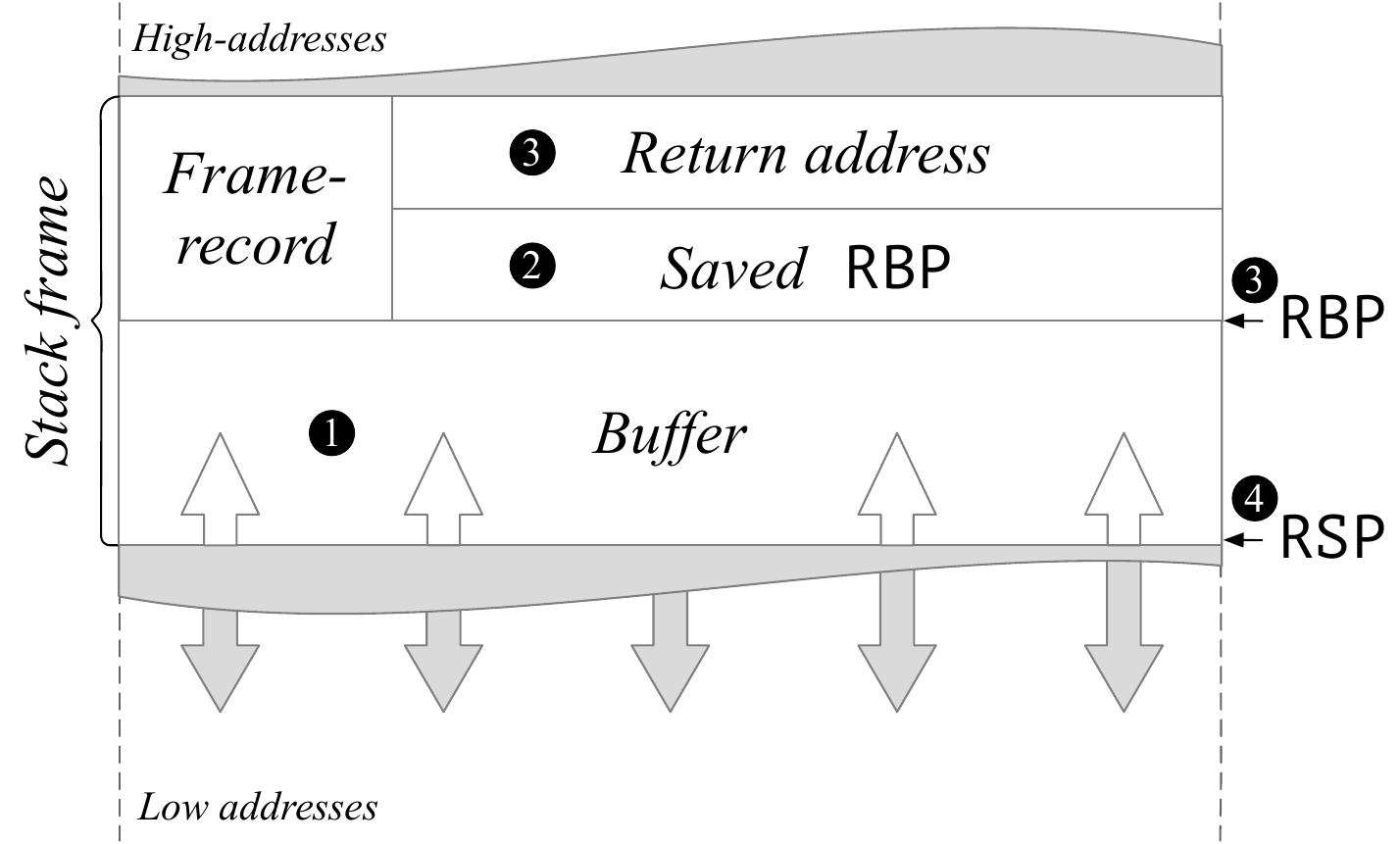}
        \caption{Stack layout without stack protection}\label{fig:nostack-protector}
    \end{subfigure}
    \begin{subfigure}[t]{0.49\textwidth}
        \includegraphics[width=\linewidth]{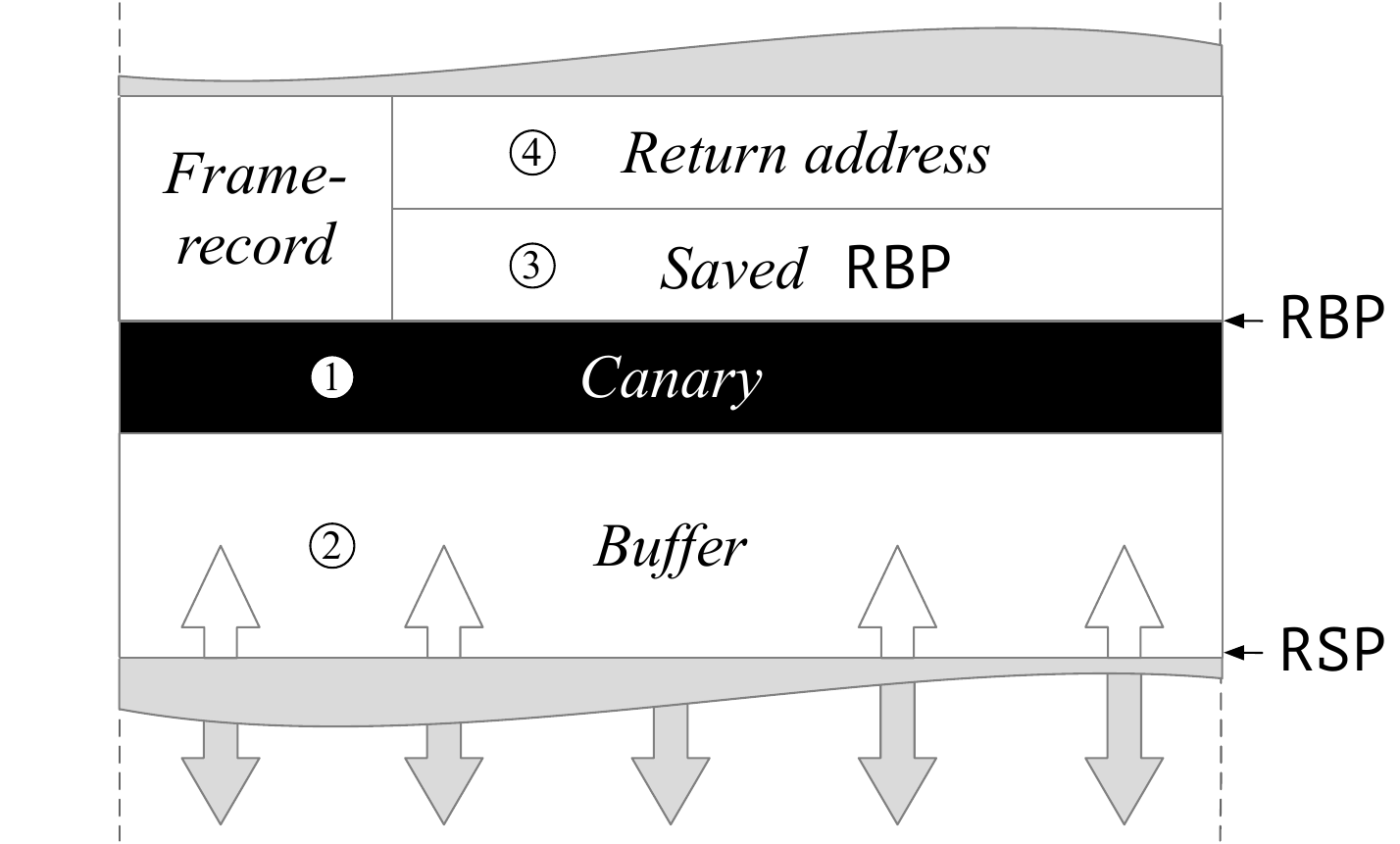}
        \caption{Stack layout with \texttt{-fstack-protector}}\label{fig:stack-protector}
    \end{subfigure}
    \begin{subfigure}[t]{0.49\textwidth}
        \includegraphics[width=\linewidth]{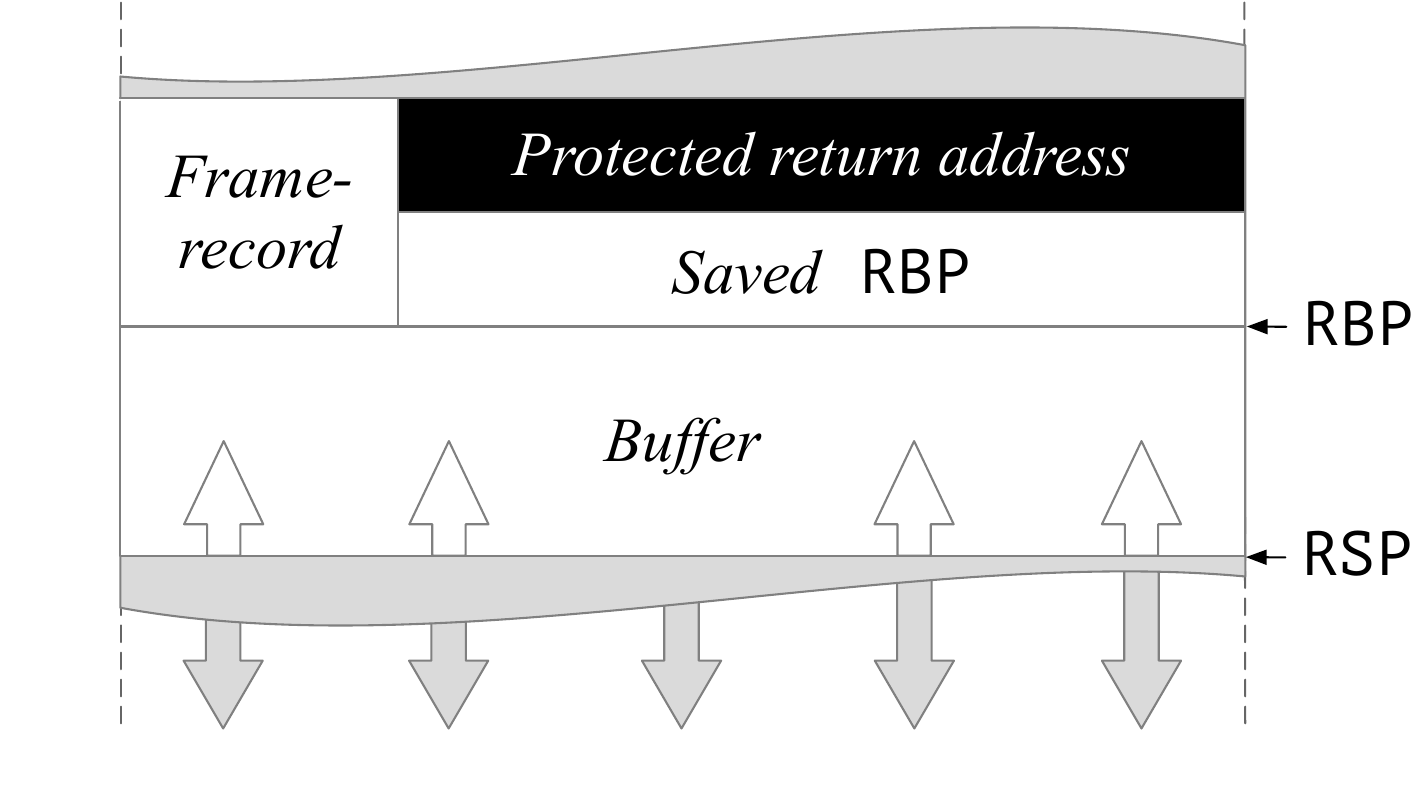}
        \caption{Stack layout with \texttt{-fcf-protection}}\label{fig:fcf-protection}
    \end{subfigure}
    \begin{subfigure}[t]{0.49\textwidth}
        \includegraphics[width=\linewidth]{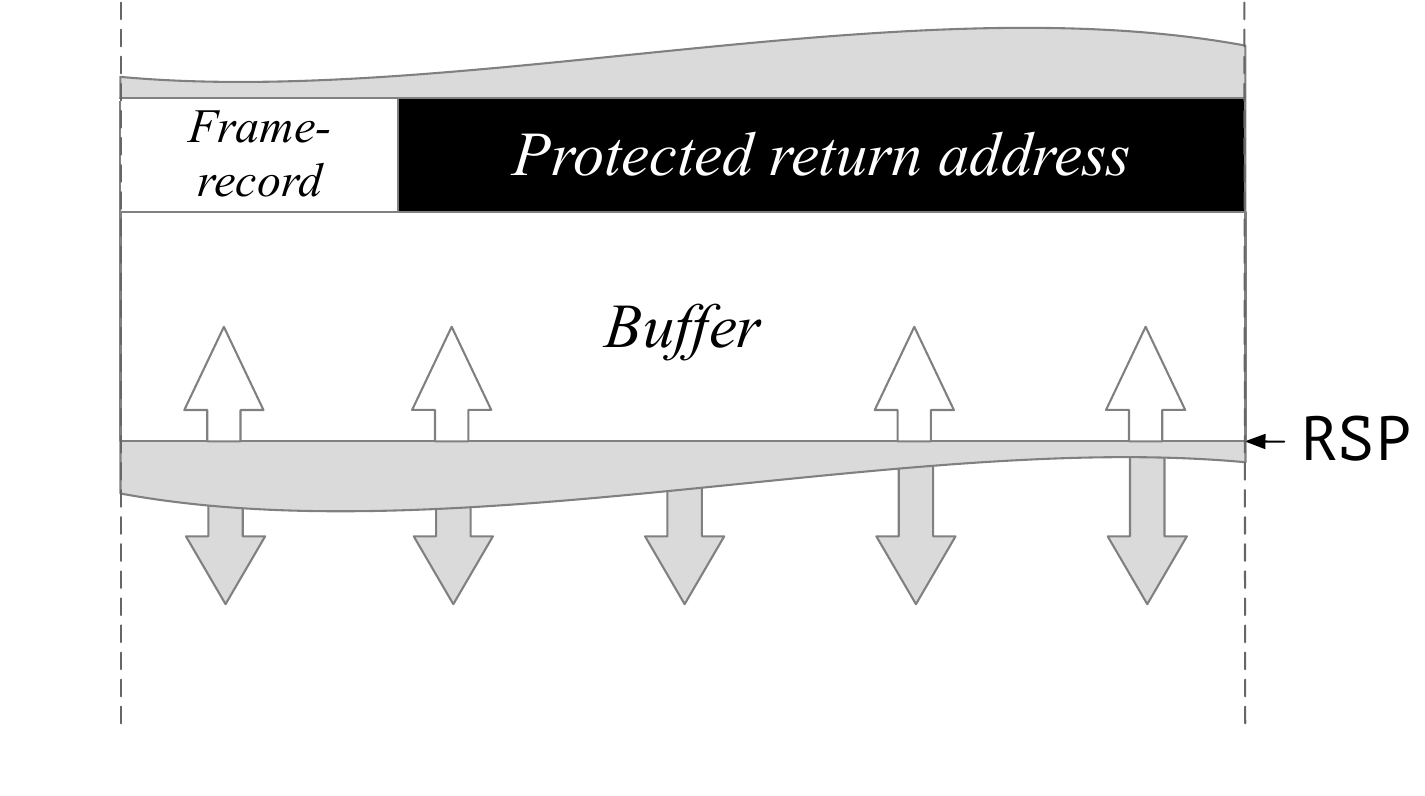}
        \caption{Stack layout with \texttt{-fcf-protection} with frame pointers omitted}\label{fig:omit-frame-pointer}
    \end{subfigure}
    \caption{Stack layout in the \gls{x86-64} architecture~(\ref{fig:nostack-protector}). A stack canary~(\ref{fig:stack-protector} \dCOne) is placed on the stack between the local variables (\dCTwo) and saved \acs{rbp} value (\dCThree) so that the canary will be overwritten in case a stack buffer overflows into the frame record. A shadow stack protects the return address in the frame record~(\ref{fig:fcf-protection}). If the frame pointer (\acs{rbp}) is omitted, the protected stack pointer acts as a canary value~(\ref{fig:omit-frame-pointer}).}\label{figure:stack-layout}
\end{figure*}

Buffer overflows are one of the first well-documented memory-safety bugs that saw a large increase in both discovered and exploited cases in the middle-90's~\cite{Levy96}.
A ''\emph{buffer}`` is simply a contiguous block of computer memory that holds multiple instances of the same data type.
In C and C++, buffers are commonly implemented as character arrays.
Arrays, like all variables in C and C++, can be declared either static or dynamic.
Static variables are allocated at load time from the executable's data or \gls{bss} segment.
In contrast, dynamic variables are allocated automatically from the program's stack, or explicitly at run time from the program's heap.
An ''\emph{overflow}'' occurs when the program erroneously writes beyond the upper bound of the allocated array.
Conversely, an ''\emph{underflow}'' occurs when writes occur beyond the lower bound of the array.
Buffer overflows can, under the right circumstances, be exploitable by memory attacks regardless of where the underlying array is allocated.
However, early exploitation techniques such as ``\emph{stack smashing}''~\cite{Levy96} target buffer overflows that occur in stack-allocated arrays since these are particularly easy to exploit.
This is because, as shown in \Cref{fig:nostack-protector}, stack-allocated buffers (\dOne) are placed just below the frame record containing the \emph{saved frame pointer} (\dTwo) and \emph{return address} (\dThree).
In the x86 \gls{isa}, including its contemporary \gls{x86-64} variant~\cite{Bendersky11}, the \glsdesc{rbp} (\acs{rbp} in \gls{x86-64} and \acs{ebp} on x86) records the beginning of local variables in function's stack frame throughout the execution of a function.
In contrast, the \acs{rsp} records the end of the stack frame.
The saved \acs{rbp} holds the value of \acs{rbp} for the caller functions, while the return address holds the address succeeding the \texttt{call} instruction in the caller that transferred control to the function.
By corrupting either the saved \acs{rbp} or return address an attacker can
override the location of the previous stack frame
(\emph{stack-pivoting}~\cite{Prakash15}) or the address the function returns to
(\emph{control-flow hijacking}~\cite{Newsome08}).

\subsection{Stack canaries}\label{sec:stackcanaries}
As the exploitation techniques for buffers overflow bugs became prevalent, research into countermeasures resulted in several mitigation schemes~\cite{Cowan98,Vendicator00,Etoh00} of which \emph{stack canaries} were eventually integrated into mainstream compilers~\cite{Whitney21,Guelton22}.
Stack canaries detect a stack buffer overflow before the execution of malicious code can occur.
\Cref{fig:stack-protector} illustrates a function's stack layout instrumented with stack canaries, where stack canaries (\dCOne) are values placed on the stack between a function's local variables (\dCTwo),  and the stored \acs{rbp} (\dCThree).
A linear buffer overflow modifies the stack canary before corrupting stored \acs{rbp} and return address (\dCFour).
A check inserted by the compiler before returning from a function detects if the canary has been modified and calls an error-handling routine, \texttt{\_\_stack\_chk\_fail()}, that typically terminates the program, rather than allowing the function to return using a corrupt return address.

\subsubsection{Canary values.} While the operating principle of stack canaries remains the same across implementations there are different approaches to choosing the canary values which affect their security against adaptive attackers:

\paragraph{Terminator canaries} include commonly used terminators such as \verb+'\0'+ or \verb+'\n'+ to prevent attackers from overflowing buffers beyond the canary without modifying it through misuses of string functions, such as \texttt{strcpy}, which copies bytes from the source until a terminator is reached.
Non-ASCII characters or invalid Unicode can be added to protect against overflows in text-only protocols.

\paragraph{Random canaries} use a random value as the canary.
The value is typically generated by the C standard library initialization code and stored in a randomized location in program memory.
However, since the reference value must be accessed to set and check it, and because the canary is stored unprotected on the stack, randomized canaries are susceptible to information leakage that reveal their value.

\paragraph{Random XOR canaries} are random canaries that are additionally XORed against a non-static value in the program (usually the \acrshort{rbp}).
In modern \glspl{os} that leverage \gls{aslr} the \acs{rbp} value for a particular function invocation will vary across different runs of the program. This adds an extra layer of randomization to the value, making it more difficult to predict.

On \gls{x86-64} Linux with the \gls{glibc}, the stack canary is a 64-bit random
value with the final bytes zeroed to make it simultaneously act as a
terminator canary.

\subsubsection{Variable placement.}
To improve stack canary efficacy, objects on the stack that are more likely to overflow should be placed closer to the stack canary so that overflows are more likely to overwrite the canary.
If there are multiple arrays in a function, arrays further from the canary could overflow into arrays closer to the canary without the overflow being detected.
Both GCC and Clang use the following rules when deciding the local variable layout~\cite{Magee17}:
\begin{itemize}
  \item Large arrays and structures containing large arrays ($\ge$ \texttt{ssp-buffer-size}) are nearest to the canary.
  \item Small arrays and structures containing small arrays ($<$ \texttt{ssp-buffer-size}) are next nearest to the canary.
  \item Variables that have had their address taken are the third nearest to the canary.
  \item Other variables are further still.
\end{itemize}

\subsubsection{Limitations.}
Stack canaries are often insufficient for stopping sophisticated attacks as they can be bypassed through exploitation primitives that corrupt memory non-sequentially.
They also may not protect against data-only attacks~\cite{Cheng19}.
Alternatively, canary values can be guessed or directly read, e.g., by repeated attempts to attack a program that does not change its stack canary after crashing, which can happen when forking, or by exploiting information leaks.
Typically, the stack canary value is the same in all functions and can, e.g., be retrieved from uninitialized stack memory from a previous function call~\cite{Hawkes06,Bittau14,Bierbaumer18}.
Nevertheless, stack canaries are routinely enabled by Linux distributions for software that they distribute.
A survey of deployed compiler-based mitigations indicates that stack canaries are enabled in $85\%$ of desktop binaries~\cite{Yu22}.

\subsection{Shadow stacks}

A shadow stack~\cite{Burow19} is a mechanism to protect a function's stored return address while it resides on the call stack.
To achieve this, a copy of the return address is stored in a separate, isolated region of memory area that is not accessible to the attacker.
Before the function returns, its stored return address is compared against the protected copy on the shadow stack to ensure the original address has not been modified, for example as a result of a buffer overflow.
If there is a mismatch between the return address on the call stack and its copy on the shadow stack, program execution is terminated.

By protecting the integrity of return addresses, shadow stacks ensure that returning from function calls leads back to the respective call site, a form of backward-edge \gls{cfi}~\cite{Abadi05}.
Attacks that violate \gls{cfi} have been demonstrated at different levels of semantic granularity, across programming languages, and in the presence of defensive mechanisms~\cite{Shacham07,Erlingsson10,Bletsch11,Roemer12,Bittau14,Hu16,Evans15,Bierbaumer18}.
The prevalence of \gls{rop}, in particular, have prompted processor manufacturers to incorporate hardware support for shadows stacks into all processor architectures including \gls{x86-64}~\cite{Corbet22}, AArch64~\cite{Corbet23}, and RISC-V~\cite{Traynor24}.
On \gls{x86-64} hardware shadow-stack support is provided by Intel's \gls{cet} as well as AMD's Shadow Stack hardware features.
At the time writing, recent releases of commodity Linux distributions, such as Ubuntu 18.04 ship with the necessary software support for \gls{x86-64} shadow stacks, but software built with shadow stack support (\texttt{-fcf-protection=return} in GCC 8.0.1 and Clang 7.0.0 and later) must explicitly opt-in to shadow-stack enforcement using a \gls{glibc} tunable (\texttt{glibc.cpu.hwcaps=SHSTK} in \gls{glibc} 2.39 and later).
Additionally, a Linux Kernel 6.6 or later built with \texttt{CONFIG\_X86\_USER\_SHADOW\_STACK=y} and a CPU with \gls{x86-64} shadow stack support (AMD Zen 3 or Intel Tiger Lake, Alder Lake, or Sapphire Rapids and later microarchitectures) is required.

The \gls{x86-64} shadow stack is stored in protected pages in a process' address space and is accessible only through specific instructions.
The \texttt{call} and \texttt{ret} instructions implicitly push, pop, and check return addresses on the call and shadow stack.
\Cref{fig:fcf-protection} illustrates the call stack layout with \texttt{-fcf-protection} enabled.
The layout is identical to \Cref{fig:nostack-protector}, but the integrity of the return address is protected by the comparison to the protected copy on the shadow stack.

\subsubsection{Comparison of stack canaries and \gls{x86-64} shadow stack.}
\Cref{tab:comparison} shows a high-level comparison between stack canaries and the \gls{x86-64} shadow stack.
The \gls{x86-64} shadow stack operates as a mechanism similar to stack canaries to protect the return address.
However, due to its placement, the \gls{x86-64} shadow stack cannot protect the
frame pointer, whereas stack canaries the detect corruption of the frame pointer
and the return address.
Stack canaries rely on heuristics to determine which functions receive the canary instrumentation based on the option shown in \Cref{tab:comparison} (with the exception of \texttt{-fstack-protector-all} which applies to all functions).
The \gls{x86-64} shadow stack applies implicitly to all functions.
Stack canaries will detect any linear and contiguous stack buffer overflow that
overwrites the canary value but can be bypassed if the canary value becomes
known to an adversary that subsequently can overflow the buffer, overwriting the
canary with its original value, or if the buffer overflow is not contiguous the
adversary can ``skip'' over the canary without overwriting it. The \gls{x86-64}
shadow stack can prevent any use of the overwritten return addresses but may be
more easily guessable than stack canaries and could allow overwriting data
further down the stack. On the other hand, the return address benefits from
\gls{aslr}, whereas stack canary entropy derives from the initial value chosen
by the glibc initialization code.
\newcommand{\notemark}[1]{\textsuperscript{#1}}
\newcommand{\tablenote}[1]{\footnotesize{#1}}

\begin{table}[ht]
    \centering
    \caption{Comparison between stack canaries and the \gls{x86-64} shadow stack}\label{tab:comparison}
    \begin{threeparttable}
    \begin{tabular}{l c c c c}
    \toprule
        \multirow{3}{*}{\qquad\quad{}Compiler option} & \multicolumn{2}{c}{Protection of frame record} & \multicolumn{2}{c}{Characteristics} \\
        \cmidrule(lr){2-3}\cmidrule(lr){4-5}
        & \splitcell{Frame \\ pointer} & \splitcell{Return \\ address} & \splitcell{Protection \\ coverage} & \splitcell{Enforcement \\ model} \\ \midrule
          \multicolumn{1}{c}{\textbf{Stack Canaries}} & & & & \\
        \rowcolor{gray!10} \texttt{-fstack-protector}        & \cmark{} & \cmark{} & Heuristic\notemark{1} & Probabilistic \\
        \rowcolor{white}   \texttt{-fstack-protector-strong} & \cmark{} & \cmark{} & Heuristic\notemark{2} & Probabilistic \\
        \rowcolor{gray!10} \texttt{-fstack-protector-all}    & \cmark{} & \cmark{} & All functions         & Probabilistic \\ \midrule
          \multicolumn{1}{c}{\textbf{Shadow stack}} & & & & \\
        \rowcolor{gray!10} \texttt{-fcf-protection=return}   & \xmark{} & \cmark{} & All functions         & Deterministic \\
    \bottomrule
    \end{tabular}
    \begin{tablenotes}
    \item[1]: \tablenote{\texttt{-fstack-protector} applies stack canaries to any function with character arrays that equal or exceed the \texttt{ssp-buffer-size} setting set via \texttt{----param=ssp-buffer-size} (8 by default).}
    \item[2]: \tablenote{\texttt{-fstack-protector-strong} applies stack canaries to any function that
        \begin{inparaenum}[1)]
            \item takes the address of any of its local variables on the right-hand-side of an assignment or as part of a function argument,
            \item allocates a local array, or a struct or union which contains an array, regardless of the type of length of the array,
            \item has explicit local register variables.
        \end{inparaenum}}
    \end{tablenotes}
    \end{threeparttable}
\end{table}

\subsubsection{Omitting the frame pointer.}

\glsreset{rbp}
\glsreset{rsp}

In \Cref{fig:nostack-protector,fig:stack-protector,fig:fcf-protection} the \gls{rbp} serves as a stable reference to the beginning of a function’s stack frame.
This is convenient in hand-written assembly and during debugging but compilers can track offsets from the \gls{rsp}, and the DWARF debugging format allows access to stack frames without a frame pointer using stored call frame information.
The System V \gls{abi} for \gls{x86-64} makes the frame pointer optional and modern compilers, including GCC and Clang, consequently omit it by default on \gls{x86-64}.
This saves two instructions in the function prologue and epilogue and makes one additional register (\acs{rbp}) available for general-purpose use.
Omitting the frame pointer precludes the need to protect the saved \acs{rbp} on the stack, resulting in the stack frame layout shown in \Cref{fig:omit-frame-pointer}.
In this configuration, barring register spills, the function's local variables are allocated immediately adjacent to the protected return address.

\section{Methodology and Challenges}\label{sec:methodology}

As we explain in \Cref{sec:background}, an often overlooked problem in validating compiler-based hardening features is test coverage and assurance of correctness.
In normal application development, the codebase is finite and known; developers focus on ensuring that all code paths within their application are tested and function correctly.
Test coverage in this setting aims to exercise as many scenarios as possible within the application’s context.
A compiler, in contrast, is used by countless developers to build a variety of applications.
When introducing a new security feature into a compiler, developers must ensure that it operates correctly across codebases, not just a single application.

In reality, most security hardening features are tested by just a small number of regression or unit tests~\cite{Beyls24}.
Even widely deployed features, such as stack canaries, can exhibit gaps
that affect their effectiveness~\cite{Hebb23} as applying them to large
amounts of code successfully does not necessarily establish their effectiveness; it just demonstrates the feature does not interfere with the normal operation of the code.
To evaluate effectiveness, a common approach is to use vulnerable programs, i.e., known \glspl{cve}.
However, \gls{cve}-based evaluation is limited both in scope, granularity, and scalability as proof-of-concept exploits are available for relatively few \glspl{cve}.
A more systematic approach is to use a benchmark suite such as RIPE~\cite{Wilander11}, Juliet~\cite{Boland12}, RIPEMB~\cite{Tauner22} or RecIPE~\cite{Jiang22}.

\subsection{Goal and problem statement}
The goal of our evaluation is to confirm or reject the following hypotheses:

\begin{itemize}
  \item[\textbf{$H_1$:}] The detection rates of stack canaries and the \gls{x86-64} shadow stack are consistent across different compilers.
  \item[\textbf{$H_2$:}] The \gls{x86-64} shadow stack has comparable effectiveness to stack canaries against linear overflows, particularly in detecting return address corruption.
  \item[\textbf{$H_3$:}] The \gls{x86-64} shadow stack exhibits better performance compared to stack canaries in real-world use cases.
\end{itemize}

To evaluate $H_1$ and $H_2$, we use the Juliet test suite~\cite{Boland12}.
It is a collection of C/C++, C\#, and Java programs with known defects organized by the corresponding \gls{cwe} categories.
The latest version released in 2017 covers 64099 C/C++ cases, 28942 C\# cases, and 28,881 Java cases.
Although the test cases in the Juliet suite are artificial, the defects in it are sourced from real-world applications, including known \glspl{cve}.
Juliet is primarily intended as a benchmark for static program analysis where the structure and syntax of a program’s code are evaluated for potential defects and vulnerabilities without executing it.
Nevertheless, the Juliet test cases are portable and self-contained, and the vast majority can compiled and run on a modern \gls{x86-64} system while exhibiting behavior that triggers memory flaws at run time.
Juliet is, to the best of our knowledge, the largest dataset of heterogeneous defective code samples available to date.
While benchmark suites such as RIPE~\cite{Wilander11} and RecIPE~\cite{Jiang22} aim to mimic attacker behavior and vary, for instance, the target memory location that is corrupted, they rely on a low number of templates that, for stack-based overflows in particular, exhibit no variation in the stack layout of surrounding the vulnerable buffer.
RIPEMB~\cite{Tauner22} is specifically geared towards embedded system evaluation.
Consequently, the Juliet C/C++ test suite is best suited for our evaluation of as it offers the largest variance in terms of test cases of the options we considered.
That said, using Juliet for run-time evaluation, rather than the static analysis it was designed for, comes with a number of challenges (\Cref{sec:challenges}).

To evaluate $H_3$, we use the SPEC CPU 2017 benchmark suite and report the
results in \Cref{sec:performance}, using \texttt{-O2 -march=native} for all
cases, with 4 copies for rate tests, and 12 threads for speed tests
(corresponding to the number of cores without \gls{smt}).
\gls{smt} and \gls{aslr} were disabled for all tests.

\subsection{Challenges}\label{sec:challenges}
We identified several challenges in using the Juliet C/C++ test suite
to evaluate the effectiveness of stack canaries and the \gls{x86-64} shadow stack for
stack-based overflows:

\subsubsection{Test case selection.} Not all of the 118 \gls{cwe} categories covered by the Juliet test suite exhibit buffer overflow defects. To keep the compilation and run-time of tests manageable, we had to narrow down the subset of test cases to evaluate those that exhibit linear buffer overflow behavior.

Through empirical assessment, we narrow our evaluation to the five \gls{cwe} categories in \Cref{tab:cwe_categories} which exhibit relevant defects.
Although our evaluation focuses on stack-based overflows, we found that category ``CWE122 Heap-Based Buffer Overflow'', despite its name, includes cases that lead to overflows in stack-based variables. The ``CWE194 Unexpected Sign Extension'' and ``CWE 195 Signed-to-Unsigned Conversion Error'' categories include tests that are relevant to our evaluation, as they involve invocations of functions such as \texttt{memcpy()} or \texttt{memmove()} with incorrect bounds.

Each test case has a ``bad'' version that exhibits a defect, and one or more ``good'' versions showing the test case with the defect patched.
We exclude the ``good'' versions from our evaluation.
Some tests have ``listen'' and ``connect'' variants that act as sources and sinks for the purposes of static analysis but are not accompanied by programs sending or receiving the required data for a fault to occur.
We chose to exclude these tests as they are not fully functional.
Finally, we exclude tests targeting the Win32 \gls{api} as we perform our evaluation in a Linux-based environment.
The \textit{Detectable} column in \Cref{tab:cwe_categories} shows the number of test cases in each category that is detectable by \emph{either} stack canaries or the \gls{x86-64} shadow stack.
We verified experimentally that the other 113 categories in the Juliet C/C++ test suite do not exhibit defects that are detectable by either of the schemes we evaluate.

\begin{table}[tb!]
	\centering
	\caption{Relevant CWE categories in Juliet C/C++ version 1.3.}\label{tab:cwe_categories}
	\resizebox{\textwidth}{!}{%
        \begin{tabular}{llrrrr}
	\toprule
	  \multicolumn{2}{c}{\textbf{CWE Category}} & \multicolumn{4}{c}{\textbf{\# Test Cases}} \\
	  \cmidrule(lr){3-6}
	  & & \textit{Total} & \textit{Excluded} & \textit{Selected} & \textit{Detectable} \\ \midrule
	  CWE121       & Stack-Based Buffer Overflow          & 4944 &  96 & 4848 &  3562 \\
	  CWE122       & Heap-Based Buffer Overflow           & 5922 & 192 & 5730 &  1426 \\
	  CWE124       & Buffer Underwrite                    & 2048 &  96 & 1952 &   604 \\
	  CWE194       & Unexpected Sign Extension            & 1152 & 384 &  768 &   192 \\
	  CWE195       & Signed-to-Unsigned Conversion Error  & 1152 & 384 &  768 &   288 \\ \bottomrule
	\end{tabular}
	}
\end{table}

\subsubsection{Test randomness.} Many Juliet test cases include an element of randomness, e.g., control-flow decisions that involve a random value, to prevent compilers from optimizing away certain code paths as dead code. Another source of randomness is \gls{aslr} employed in modern \glspl{os}, which vary both the location of program code and data segments in memory as well as the starting address of the stack and heap. For our results to be reproducible we had to disable the sources of nondeterminism in the tests, while ensuring that the compiler does not leverage optimizations that benefit from dead code removal or certain types of undefined behavior that the test cases exhibit by design.

To eliminate randomness in tests that use \texttt{srand(time(NULL));} to set a random seed (sometimes multiple times in the program) and the \texttt{rand()} function to obtain random values used by the test we used wrapper library interposed through \texttt{LD\_PRELOAD} to intercept \texttt{srand(time(NULL));} and set a fixed seed value for the test.
We also disabled \gls{aslr} via the \texttt{/proc/sys/kernel/randomize\_va\_space} kernel interface.
A drawback of this approach is that it precludes alternate possible behaviors in some tests but makes the test results consistently reproducible across runs.
We also used \texttt{env ----ignore-environment} in our tests to prevent the influence of environment variables, such as SWAYSOCK (specific to our graphical interface) which we have seen to affect test results.

Because stack canary values are randomized by startup code part of glibc itself,
our approach of interposing glibc does not affect stack canary initialization.
We found that, in some test cases, the canary value itself is misinterpreted as an address as a result of the overflow, the program to crashes in either a segmentation fault (\texttt{SIGSEGV}) or a bus error (\texttt{SIGBUS}) depending on the value of the stack canary.
This shortcoming could be addressed by patching glibc to initialize stack canaries with a fixed value, but we chose to leave the behavior as is and interpret both signals similarly, as valid values should be rare, and both cases correspond to a detection failure in our experimental setup.

\subsubsection{Discerning test outcomes.} Since both successful detection and failure to detect a buffer overflow are likely to lead to the test program crashing we need to discern between the different outcomes to be able to attribute the crash to either a stack canary check, \gls{x86-64} check, or other program crash.
In order to determine the reason for the crash, we examine
\begin{inparaenum}[1)]
  \item the program output to detect whether it exhibits \texttt{*** stack smashing detected ***: terminated}, which is generated by glibc’s \texttt{\_\_stack\_chk\_fail} function when stack-canary check fails,
  \item the program trace generated using \texttt{strace} to attribute segmentation faults with \texttt{si\_code=SEGV\_CPERR} and \texttt{si\_addr=NULL} to shadow stack violations, or
  \item a zero or non-zero exit status and terminations due to \texttt{SIGSEGV}, \texttt{SIGBUS}, \texttt{SIGILL} or \texttt{SIGFPE} or as detection failures.
\end{inparaenum}

\subsubsection{Sensitivity to compiler optimizations and options.} Due to the self-contained nature and small size of many Juliet test cases, they are sensitive to various compiler optimizations that can alter the stack layout, e.g., change the order of stack variables, or avoid placing certain variables on the stack altogether. The optimization strategies between GCC and Clang are not identical, so to account for this variance, we need to repeat our experiments under different optimization levels across GCC and Clang.

To account for these variations we ran all tests with multiple compilation options.
More specifically, we tested differences between the \texttt{--O2}, and \texttt{--O0} optimizations levels to ensure weaknesses are not simply optimized over, and multiple heuristics for stack canaries, shown in \Cref{tab:comparison}: stack canaries disabled, stack canaries in every function, and \verb+-fstack-protector-strong+ and \verb+-fstack-protector+ with values of 4 and 8 for \verb+ssp-buffer-size+.

\subsection{Experimental setup}

We opted to use a source-based Linux distribution, in this case Gentoo Linux, to ensure that not only the test cases themselves, but also all dependencies were built with the corresponding stack canary and \gls{x86-64} shadow stack options and the correct compiler.\footnote{\texttt{COMMON\_FLAGS} is set to \texttt{-O2 -pipe -march=native -fcf-protection}, profile is \texttt{default/linux/amd64/23.0}.}
We used GCC 13.3.1\_p20240614 p1 and Clang version 18.1.8 along with Gentoo's glibc 2.39-r6, on Gentoo’s Linux Kernel version 6.6.51-gentoo-dist-hardened (system updated on September 30 2024, 8:41 UTC).
We used a Intel NUC 13 Pro Mini (NUC13ANK) with 13th Generation Raptor Lake Intel Core™ i7-1360P processor and 14~GB \gls{ram} available to Gentoo Linux for our experiments.

\section{Evaluation and Results}

In this section, we report on the results of the experiments outlined in \Cref{sec:methodology}.

\subsection{Results: Detection of Linear Overflows}\label{sec:detection}

\newcommand{\mydef}[2]{
  \bigbreak
  \begin{tcolorbox}[boxrule=0.2mm,title={\textbf{#1}}]
    \textit{{#2}}
  \end{tcolorbox}
  \vspace*{-0.75cm}
  \bigbreak}

\Cref{fig:bad_tests} illustrates the stack canary and shadow stack results for GCC and Clang under the optimization levels \texttt{--O2} and \texttt{--O0}.
We focus on “bad” versions of Juliet’s test cases as the corrected versions showed no false positives for either stack canaries or the shadow stack.
All test cases show better detection rates with \texttt{--O0} compared to \texttt{--O2}, as the vulnerable code portions are often optimized away under \texttt{--O2}.
Consequently, for comparison, it is more meaningful to compare the detection rates \emph{within} a certain optimization level than the rates \emph{across} optimization levels.

\mydef{Stack canary detection rates}{Clang demonstrates better detection rate with stack canaries than GCC.}

\subsubsection{Stack canary detection rates.}

An overall comparison of the plots in \Cref{fig:bad_tests} reveals that Clang demonstrates better detection rate with stack canaries than GCC\@.
The \texttt{--fstack-protector-all} and \texttt{--fstack-protector-strong} options consistently outperform the \texttt{--fstack-protector} option, which is expected.
The \texttt{--fstack-protector\\-all} option does not perform significantly better than \texttt{--fstack-protector-strong}.
There is a slight difference in favor of \texttt{--fstack-protector-all} in the GCC~\texttt{-O0} case.
The number of tests with a stack canary detection with \texttt{-fno-stack-protector} is nonzero, because some C library functions, such as \texttt{memcpy}, have stack canaries on the system, no matter the compilation options used for the test (there are 8 such GCC tests, and 6 Clang tests).
We attribute the differences in detection results between GCC and Clang as follows below:

\begin{sidewaysfigure}
	\centering
	\begin{subfigure}{\textwidth}
		\centering
		\includegraphics[width=\linewidth,keepaspectratio]{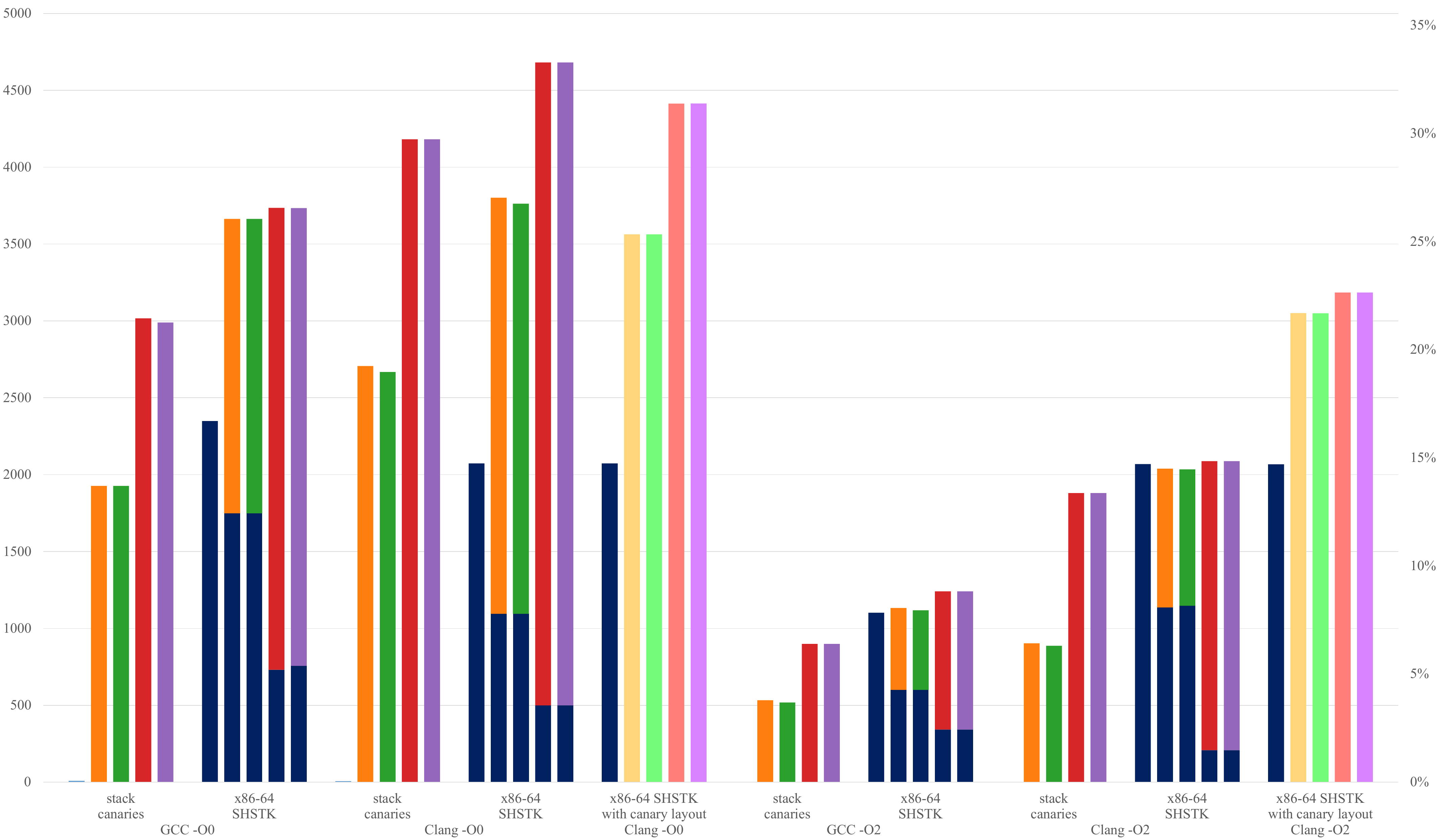}%
		\label{fig:bad_tests-gcc-O2}
	\end{subfigure}
	\begin{subfigure}[T]{0.38\textwidth}
		\centering
		\includegraphics[width=\textwidth,height=\textheight,keepaspectratio]{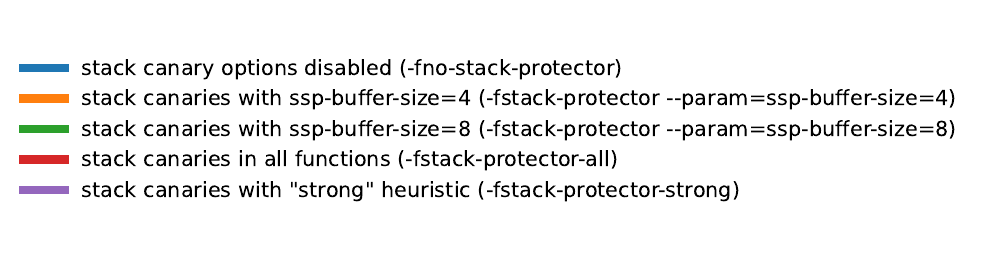}
	\end{subfigure}
	\begin{subfigure}[T]{0.49\textwidth}
		\centering
		\includegraphics[width=\textwidth,height=\textheight,keepaspectratio]{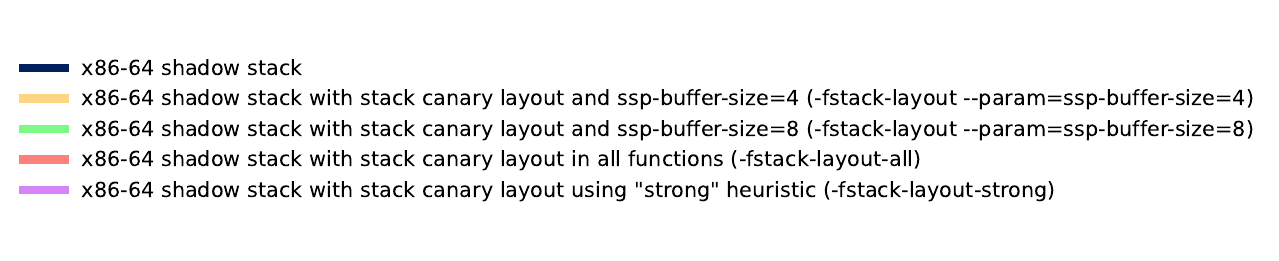}
	\end{subfigure}
	\caption{Comparison of Juliet test results by compiler, optimization level, and options. The \emph{stack canaries} bars show the detection rates for different stack canary options (indicated in the legend) with the \gls{x86-64} shadow stack disabled. The \emph{\gls{x86-64} SHSTK} bars show the detection rate for the \gls{x86-64} shadow stack separately, and when combined with different stack canary options (indicated in the legend). The \emph{\gls{x86-64} SHSTK with canary layout} bars show the detection rate for the proof-of-concept options discussed in \Cref{sec:improvedetection} for Clang \texttt{-O0} and \texttt{-O2} configurations. The left axis shows the number of test cases with detections, while the right axis shows the percentage of detected cases relative to the \emph{Selected} test cases shown in \Cref{tab:cwe_categories}. Raw values are available in \Cref{tab:juliet-detections-base} and \Cref{tab:juliet-detections-mod} in the Appendix.}\label{fig:bad_tests}
\end{sidewaysfigure}

\paragraph{Differences in stack layout between compilers:} In GCC, an array may be placed before another array, while in Clang, the same array may be placed after. This difference in stack layout can result in arrays being positioned closer to the stack canary and return addresses, depending on the compiler.

\paragraph{Differences in handling of \texttt{alloca()} calls with constant values:} Clang treats \texttt{alloca} calls with constant values similarly to a local array declaration, optimizing the allocation accordingly. In contrast, GCC employs a dynamic implementation, which may allocate additional space, particularly at the \texttt{--O0} optimization level. This behavior can allow a buffer to overflow with a specific length without modifying the stack canary.

\paragraph{Differences in stack canary options:} With the options providing larger function coverage for stack canaries outperforming the weaker heuristics.

\subsubsection{\gls{x86-64} shadow stack detection rates.}

When the \gls{x86-64} shadow stack is enabled without stack canaries present, its detection rates exceeds those of stack canaries in the \texttt{--O2} case for both GCC and Clang, but the \texttt{--O0} results are reversed with the best performing stack canary options (\texttt{-fstack-protector-all} and \texttt{-fstack-protector-strong}) detecting more overflows than the \gls{x86-64} shadow stack for both GCC and Clang.
Considering all compilation options, there are 1217 tests that the \gls{x86-64} shadow stack detects successfully that stack canaries do not, and 163 tests that stack canaries detect that the \gls{x86-64} shadow stack does not.

\mydef{\gls{x86-64} shadow stack detection rates}{The \gls{x86-64} shadow stack does not consistently outperform stack canaries.}

\begin{table}[t]
    \centering
    \caption{Geometric mean of performance degradation based on SPEC CPU
2017 results presented in \Cref{fig:spec-results-i3}.}%
    \label{tbl:spec-results-i3}
    \resizebox{\textwidth}{!}{%
    \begin{tabular}{lrr}
    \toprule
        \multicolumn{1}{c}{\textbf{Protection variant}} &
        \multicolumn{1}{c}{\textbf{intrate}} & \multicolumn{1}{c}{\textbf{intspeed}}\\
    \midrule
        stack canaries with ``strong'' heuristic
        (\texttt{-fstack-protector-strong})                                    &
        0.09\% & 0.42\% \\
        stack canaries in all functions (\texttt{-fstack-protector-all})
        & 2.18\% & 3.21\%\\
        \gls{x86-64} shadow stack
        & 0.99\% & 0.40\% \\
        \gls{x86-64} shadow stack with stack canary layout using ``strong''
        heuristic (\texttt{-fstack-layout-strong})  & 1.16\% & 0.34\% \\
        \gls{x86-64} shadow stack with stack canary layout in all functions
        (\texttt{-fstack-layout-all})               & 0.80\% & 0.25\%  \\
    \bottomrule
    \end{tabular}}
\end{table}

\begin{figure*}[t!]
    \centering
        \begin{subfigure}{\textwidth}
            \includegraphics[width=\textwidth]{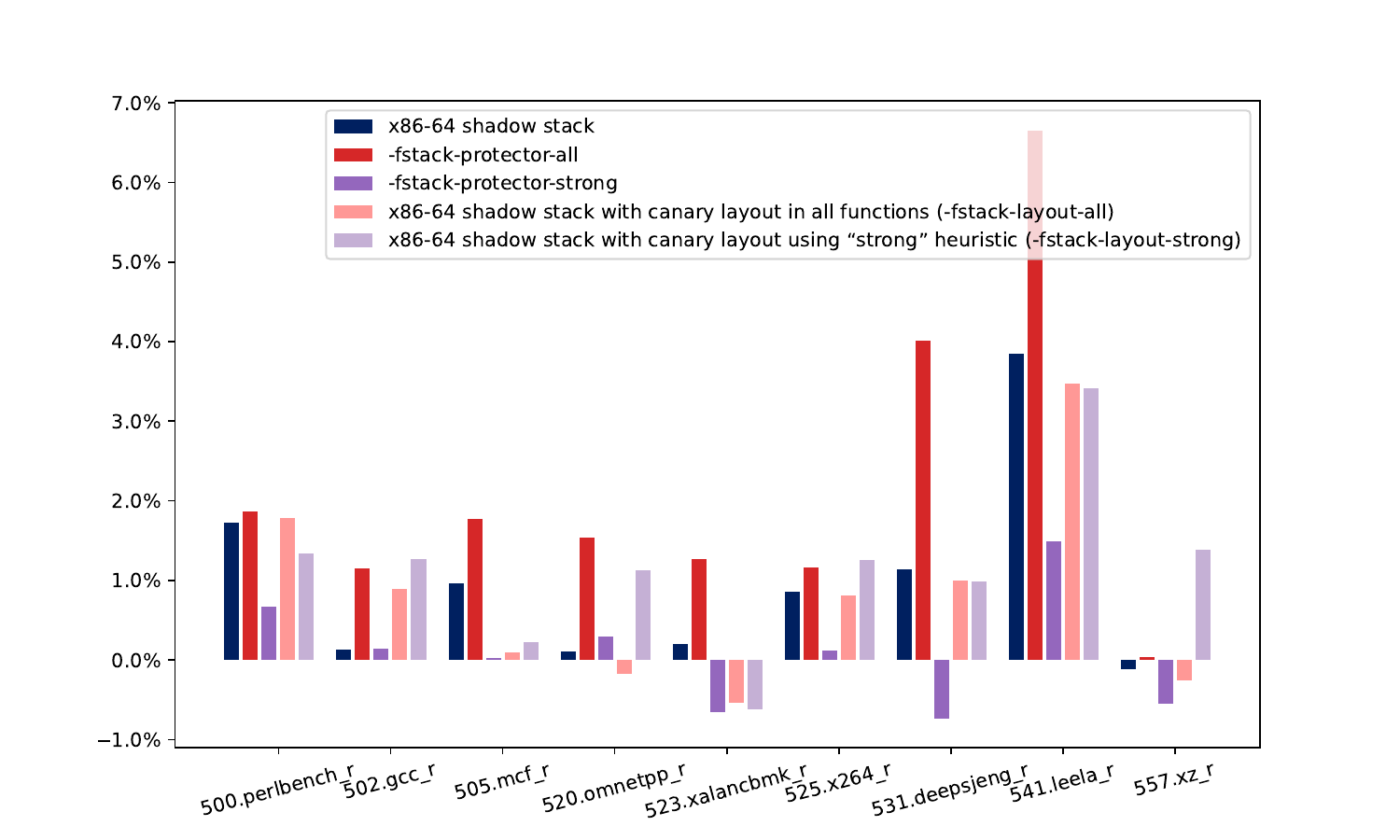}
            \caption{Results on the intrate benchmarks.}
        \end{subfigure}
        \begin{subfigure}{\textwidth}
            \includegraphics[width=\textwidth]{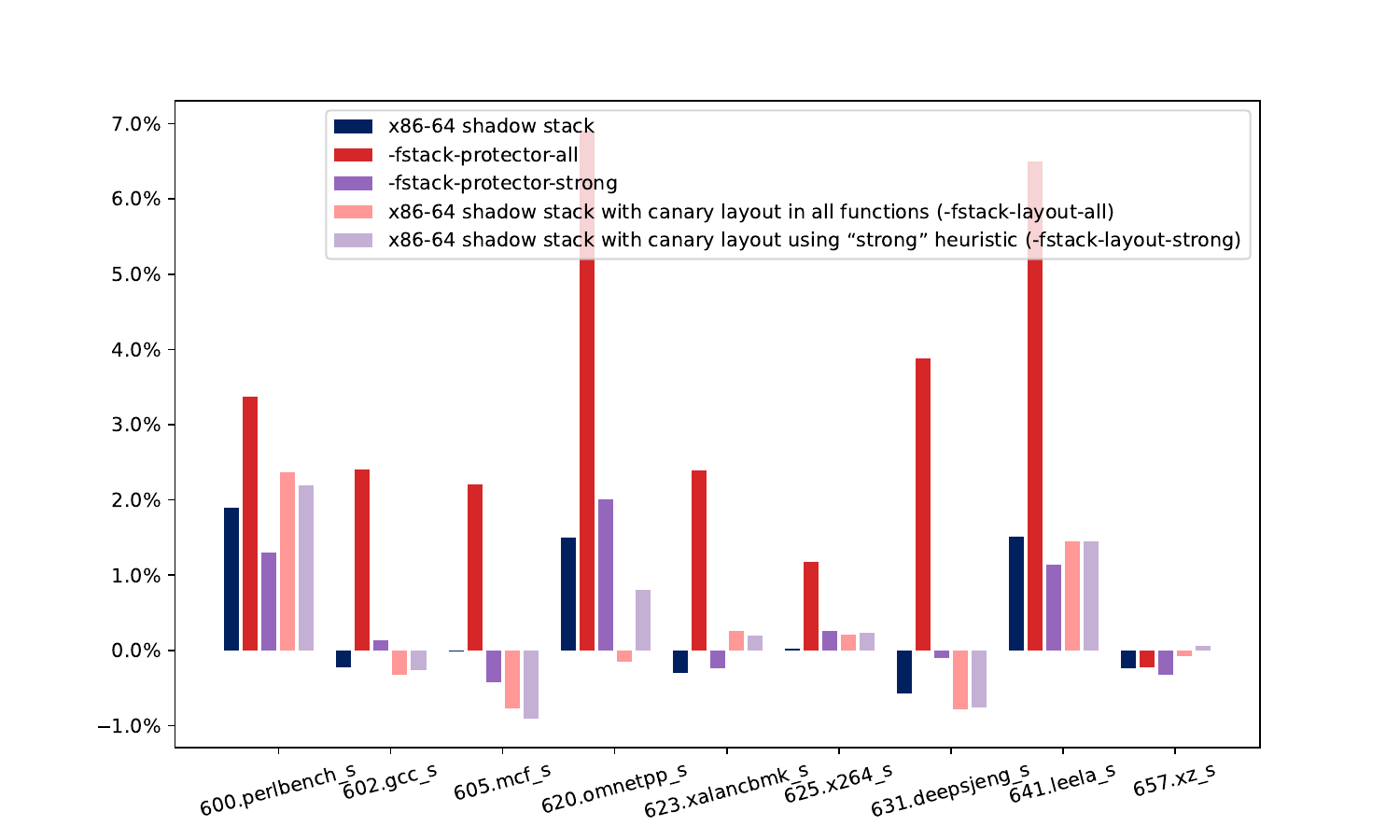}
            \caption{Results on the intspeed benchmarks.}
        \end{subfigure}
        \caption{Relative performance degradation for SPEC CPU 2017 benchmarks normalized to the baseline with \gls{x86-64} shadow stack and stack canaries disabled. The bars for the \texttt{-fstack-layout} -family of compiler flags show the performance for the proof-of-concept \gls{x86-64} shadow stack instrumentation discussed in \Cref{sec:improvedetection}.}\label{fig:spec-results-i3}
\end{figure*}

\subsection{Results: Performance}\label{sec:performance}
We evaluate the performance impact of different stack canary implementations and the \gls{x86-64} shadow stack using the SPEC CPU 2017 intrate and intspeed benchmarks.
For the performance evaluation we focus on the \texttt{-fstack-protector-strong} and \texttt{-fstack-protector-all} options as these outperformed the other \texttt{-fstack-protector} variants in the detection of linear overflows experiments (\Cref{sec:detection}).
To improve the consistency of result, we disabled \gls{aslr} and \gls{smt}.
All benchmarks were compiled using Clang compiler with optimization level~\texttt{--O2} and \texttt{-march=native}.
We exclude the \texttt{548.exchange2\_r} benchmark as it is written in Fortran and not supported by Clang.

\Cref{tbl:spec-results-i3} gives an overview of the performance
results and \Cref{fig:spec-results-i3} shows the relative
performance degradation introduced by the different options on individual
intspeed and intrate benchmarks.
Overall we found that the \texttt{-fstack-protector-strong} options degraded performance the least ($\approx$~0.09\% on rate and $\approx$~0.42\% on speed) and \texttt{-fstack-protector-all} the most ($\approx$~2.18\% on rate and $\approx$~3.21\% on speed).
The \gls{x86-64} shadow stack falls between these stack canary variants by
degrading performance by $\approx$~0.99\% on rate and $\approx$~0.40\% on speed.

\subsection{Conclusions from Evaluation}

In \Cref{sec:methodology} we set out to confirm or reject our hypotheses:

\begin{itemize}
  \item[\textbf{$H_1$:}] The detection rates of stack canaries and the \gls{x86-64} shadow stack are consistent across different compilers.
  \item[\textbf{$H_2$:}] The \gls{x86-64} shadow stack has comparable effectiveness to stack canaries against linear overflows, particularly in detecting return address corruption.
  \item[\textbf{$H_3$:}] The \gls{x86-64} shadow stack exhibits better performance compared to stack canaries in real-world use cases.
\end{itemize}

Our conclusions, based on the above evaluation of detection and performance is:

\begin{itemize}
    \item[\textbf{$H_1$:}] Rejected~\xmark{} Our results show that different options exhibit different detection rates across compilers.
    \item[\textbf{$H_2$:}] Rejected~\xmark{} The \gls{x86-64} shadow stack does not consistently outperform stack canaries in terms of detection rates.
    \item[\textbf{$H_3$:}] Rejected~\xmark{} We measured consistently larger performance impacts for the \gls{x86-64} shadow stack compared to \texttt{-fstack-protector-strong} in our benchmarks.
\end{itemize}

This outcome prompted us to consider whether we could augment the \gls{x86-64}
shadow stack instrumentation to improve its detection accuracy based on the
observation that the substantial differences in overflow detection with
stack canaries between compilers are due to differences
in stack allocation layout.

\section{Improving the \gls{x86-64} shadow stack detection accuracy}\label{sec:improvedetection}

To improve the \gls{x86-64} shadow stack detection accuracy, we implemented a modification to the Clang compiler that apply the stack layouts changes implied by the \texttt{-fstack-protector} family of options without enabling the stack canary instrumentation and checks.
To achieve this, we reuse the analyses passes that the stack canary instrumentation uses, but remove the generation of the failure path, check, and stack canary allocation.
These simple changes result in a \texttt{-fstack-layout} -family of options that make local allocations ordered by the rules described in \Cref{sec:stackcanaries}, with large arrays and structures containing large arrays closer to the return address than small arrays and variables.
We re-evaluate \gls{x86-64} shadow stack detection accuracy when combined with the new \texttt{-fstack-layout} -family of options.
The results for Clang with \texttt{--O0} and \texttt{--O2} are shown in~\Cref{fig:bad_tests}.
They show a consistent improvement in \gls{x86-64} shadow stack detection accuracy when combined with the new options.
We also repeated the performance benchmarks using the \gls{x86-64} shadow stack when combined with \texttt{-fstack-layout-strong} and report to result in \Cref{tbl:spec-results-i3} and \Cref{fig:spec-results-i3}.
We found the \gls{x86-64} shadow stack with \texttt{-fstack-layout-strong} and \texttt{-fstack-layout-all} to have comparable performance to that the conventional \gls{x86-64} shadow stack.

\subsubsection{Limitations} In some cases, such as when a function spills callee-saved registers\footnote{\texttt{\%rbx}, \gls{rsp}, \gls{rbp}, and \texttt{r12--r15} in the System V ABI~\cite{Lu18}}
 the stack canary is not placed right next to the return address, but also in such a way to protect any spilled register values.
An option that would protect these better when the return address itself is used as a “stack canary” would be to consider these as low-risk variables, and placed further from the return address than any other variable.
They would still be protected from overruns of any variable, but might be considered less protected against underflows compared to the usual stack canary case.
Our current proof-of-concept implementation of the \texttt{-fstack-layout} does not alter the placement of such spilled registers.

\section{Related Work}
The \gls{x86-64} shadow stack, particularly Intel's realization part of \gls{cet} has seen extensive evaluation focusing on its use in \gls{cfi} enforcement and its performance overhead~\cite{Shanbhogue19,Burow18,Kucab22}.
To the best of our knowledge, we are the first to evaluate its effectiveness in replacing conventional stack canaries.

Bierbaumer et al.~\cite{Bierbaumer18} conduct a thorough evaluation of stack canary security and its relation to shadow stacks.
They, however, focus on the safe stack~\cite{Kuznetzov18}, which is only
available in Clang and has known compatibility issues and limitations with
garbage collection, signal handling, and shared
libraries\cite{OpenSSFcontributors24}.

Alternatives for stack canaries have been proposed before: PCan~\cite{Liljestrand19} proposes an Arm \gls{pa}-based canary design that employs multiple function-specific canaries.
\gls{pa} is a hardware extension to the Armv8 and Armv9 \gls{isa} that principally provides backward-edge \gls{cfi}, similar to shadow stacks, but operate by embedding short, hardware-generated message authentication codes, referred to as \glspl{pac} to return addresses.
The PCan proposal, however, overlooks the impact of variable reordering, which we show has a significant impact on the effectiveness of stack canaries.

The Clang community is considering proposals for BOLT-based binary analysis tools for evaluating the effectiveness of compiler-based security hardening at the binary level~\cite{Beyls24}.
BOLT~\cite{Panchenko19} is a post-link optimizer built on top of the LLVM framework that utilizes sample-based profiling, principally for performance improvement, but has shown out to be extensible for different types of binary analysis use cases as well.
The BOLT-based analysis currently supports backward-edge \gls{cfi} provided through Arm \gls{pa} and stack-clash protection~\cite{OpenSSFcontributors24} in Clang.
Binary analysis of hardening features is orthogonal to our work and is focused on validating that the instrumentation is added correctly and consistently by the compiler to the examined binaries.
Validation using binary analysis will also benefit greatly from being applied to a large corpus of test cases, which is the focus of this paper.

\subsubsection{Similar hardware functionality in other ISAs}

Hardware-assisted shadow stacks have, in the last decade, become commonplace in several general-purpose computer architectures.
The Armv9 architecture also supports, besides \gls{pa} for backward-edge \gls{cfi}, a shadow stack implementation referred to as \gls{gcs}~\cite{Corbet23}.
The RISC-V architecture has recently incorporated support for hardware shadow stacks through the \emph{zicfiss} extensions~\cite{Traynor24}.
Each of these features could be considered a candidate for replacing stack canaries.
In future work, we plan to evaluate Armv8 and Armv9 \gls{pa}, extending our analysis to use BOLT-based \gls{pa} analysis as an additional metric.

\section{Conclusion}

This research explores the hypothesis that stack canaries and the x86
shadow stack are comparably effective for the detection of linear overflows
and return-address corruption across compilers and optimization levels,
while we expected the hardware-supported shadow stack to show better
runtime performance than stack canaries.
We use the Juliet test suite with the GCC and Clang compilers and the SPEC CPU 2017 benchmarks to test these
hypotheses. We discover that, regarding stack canaries, compiling with
Clang results in substantially better detection rates than compiling with
GCC, and that detection is generally better when compiling without
optimizations \texttt{--O0} in comparison to the commonly used \texttt{--O2} optimization
level. For \texttt{--O2}, the shadow stack generally allows for more overflows to be
detected than stack canaries; while again Clang outperforms GCC in
detection rates.
We believe that our findings should be investigated further by the
communities maintaining Clang and GCC\@. Specifically, the
lower-than-expected detection rates when using GCC could be an indication
for one compiler being able to choose a better stack layout but may also be
the result of biases in the test suite.
Our experiments indicate that \gls{x86-64} shadow stack is effective in
catching programming errors and we experimented with improving this
effectiveness by using a stack layout normally generated for stack canaries
in Clang, but without executing regular stack canary checks.
We evaluate our Clang modifications with the Juliet
test suite and observe detection rates substantially above those of
stack-canary implementations. The performance impact of the modifications to the shadow stack
seem negligible, which could make this configuration a preferable
alternative to stack canaries in some use cases and on supported hardware,
albeit with the limitation of return addresses being more
\enquote{guessable} than random stack canaries. We acknowledge that our
evaluation does not enable us to make strong claims regarding the security
of the different configurations as the Juliet test suite is not designed to
test exploitation methods.
Our results further demonstrate that stack-protector implementations have
impact on code generation beyond inserting a canary value in the function
prologue and epilogue. They also influence stack ordering between different
arrays and impact the overall stack layout. This insight suggests that
similar considerations should be applied to other return address-protection
mechanisms, such as Arm \gls{pa}, to ensure comprehensive coverage and
security.

\section*{Acknowledgements}

We want to thank Kristof Beyls, William Huhn, Siddhesh Poyarekar and Niklas Lindskog for reviewing an earlier version of this paper.
This research is partially funded by the CyberExcellence programme of the Walloon Region, Belgium (grant 2110186).

\bibliographystyle{splncs04}
\bibliography{main,local}

\newpage
\section*{Appendix}

\newcommand{\tworow}[1]{\multirow{-2}{*}{#1}}
\newcommand{\onerow}[1]{\multirow{1}{*}{#1}}

\begin{table*}[h!]
    \centering
    \caption{Number of Juliet test cases with detections by stack canaries (STKCNRY) or the \gls{x86-64} shadow stack (SHSTK), for both for GCC and Clang with optimization levels \texttt{--O0} and \texttt{--O2}. This is represented graphically in \Cref{fig:bad_tests}. Note that testing for detection by stack canaries and the shadow stack was done separately, as such, we interpreted tests detected by both (at most 12 for a given configuration) as stack canaries triggering after the shadow stack, as stack canaries aren’t disabled for shadow stacks test, while the opposite is true.}\label{tab:juliet-detections-base}
    \resizebox{\textwidth}{!}{%
    \begin{tabular}{l cc cc cc cc}\toprule
        & \multicolumn{4}{c}{\textbf{GCC}} & \multicolumn{4}{c}{\textbf{Clang}} \\
        \cmidrule(lr){2-5}\cmidrule(lr){6-9}
        & \multicolumn{2}{c}{\texttt{-O0}} & \multicolumn{2}{c}{\texttt{-O2}} & \multicolumn{2}{c}{\texttt{-O0}} & \multicolumn{2}{c}{\texttt{-O2}} \\
        \cmidrule(lr){2-3}\cmidrule(lr){4-5}\cmidrule(lr){6-7}\cmidrule(lr){8-9}
        \rowcolor{white}Detected by & \rotatebox[origin=l]{90}{STKCNRY} & \rotatebox[origin=l]{90}{SHSTK}
                                    & \rotatebox[origin=l]{90}{STKCNRY} & \rotatebox[origin=l]{90}{SHSTK}
                                    & \rotatebox[origin=l]{90}{STKCNRY} & \rotatebox[origin=l]{90}{SHSTK}
                                    & \rotatebox[origin=l]{90}{STKCNRY} & \rotatebox[origin=l]{90}{SHSTK} \\ \midrule
        \rowcolor{white}   \multicolumn{1}{c}{Stack canaries} \\
        \rowcolor{white}   \texttt{-fno-stack-protector}                         &    8 &  --  &   0 &  --  &    6 &  --  &    0 &  --  \\
        \rowcolor{gray!10} \texttt{-stack-protector ----param=ssp-buffer-size=4} & 1927 &  --  & 533 &  --  & 2707 &  --  &  903 &  --  \\
        \rowcolor{white}   \texttt{-stack-protector ----param=ssp-buffer-size=8} & 1927 &  --  & 518 &  --  & 2668 &  --  &  887 &  --  \\
        \rowcolor{gray!10} \texttt{-fstack-protector-all}                        & 3017 &  --  & 899 &  --  & 4182 &  --  & 1880 &  --  \\
        \rowcolor{white}   \texttt{-fstack-protector-strong}                     & 2991 &  --  & 899 &  --  & 4182 &  --  & 1880 &  --  \\ \midrule
        \rowcolor{white}   \multicolumn{1}{c}{\gls{x86-64} shadow stack} \\
        \rowcolor{white}   \texttt{-fno-stack-protector}                         &   0  & 2349 &   0 & 1102 &    0 & 2073 &    0 & 2069 \\
        \rowcolor{gray!10} \texttt{-fstack-protector ----param=ssp-buffer-size=4}& 1915 & 1749 & 533 &  600 & 2707 & 1095 &  903 & 1136 \\
        \rowcolor{white}   \texttt{-fstack-protector ----param=ssp-buffer-size=8}& 1915 & 1749 & 518 &  600 & 2668 & 1095 &  887 & 1147 \\
        \rowcolor{gray!10} \texttt{-fstack-protector-all}                        & 3005 &  731 & 899 &  342 & 4182 &  499 & 1880 &  208 \\
        \rowcolor{white}   \texttt{-fstack-protector-strong}                     & 2979 &  761 & 899 &  342 & 4182 &  499 & 1880 &  208 \\ \bottomrule
    \end{tabular}}
\end{table*}

\begin{table*}[h!]
    \centering
    \caption{Number of Juliet test cases with detections by the \gls{x86-64} shadow stack on modified Clang with optimisation levels \texttt{--O0} and \texttt{--O2} and different stack layout options. This is represented graphically in \Cref{fig:bad_tests}.}\label{tab:juliet-detections-mod}
    \begin{tabular}{l rr}\toprule
        \gls{x86-64} shadow stack (modified Clang) detections & \texttt{-O0} & \texttt{-O2} \\ \midrule
        \rowcolor{gray!10} \texttt{-fstack-layout ----param=ssp-buffer-size=4}   & 2067 & 3051 \\
        \rowcolor{white}   \texttt{-fstack-layout ----param=ssp-buffer-size=8}   & 3563 & 3059 \\
        \rowcolor{gray!10} \texttt{-fstack-layout-all}                           & 4414 & 3185 \\
        \rowcolor{white}   \texttt{-fstack-layout-strong}                        & 4114 & 3185 \\ \bottomrule
    \end{tabular}
\end{table*}

\end{document}